%% file: main.tex
\documentclass[runningheads]{llncs}
\usepackage[T1]{fontenc}
\usepackage{graphicx}
\usepackage{amssymb} 
\usepackage{amsmath} 
\usepackage{booktabs} 
\usepackage{siunitx}
\usepackage{hyperref}
\usepackage{comment}
\usepackage{makecell} % put in your preamble
\usepackage{todonotes}

\usepackage{xcolor}

\begin{document}

%\title{The Last MEV Bite:\\Sandwich Attacks in Private L2 Mempools}
\title{How to Serve Your Sandwich?\\MEV Attacks in Private L2 Mempools}

\titlerunning{Sandwich Attacks in Private Mempools} 

\author{Krzysztof M. Gogol\inst{1,2} \and Manvir Schneider\inst{3} \and Jan Gorzny \inst{2} \and \\ Claudio J. Tessone\inst{1,4}}
\authorrunning{K. Gogol, M. Schneider, J. Gorzny, C. Tessone}

\institute{University of Zurich UZH \and Zircuit \and Cardano Foundation \and UZH Blockchain Center}

\maketitle              % typeset the header of the contribution
\begin{abstract}

\input{sections/00Abstract.tex}
\end{abstract}

\input{sections/01Introduction}
%\input{sections/02Background}
\input{sections/03Model_v3}

\input{sections/04Enforcing-order}

\input{sections/05Methodology}
\input{sections/06EmpiricalResults}

\input{sections/08ConclusionDiscussion}

%\newpage
\bibliographystyle{splncs04}
\bibliography{main}
	
\appendix
\input{sections/96Appendix}

\end{document}

%% file: sections/00abstract.tex
We study the feasibility, profitability, and prevalence of sandwich attacks on Ethereum rollups with private mempools. 
First, we extend a formal model of optimal front- and back-run sizing, relating attack profitability to victim trade volume, liquidity depth, and slippage bounds. We complement it with an execution-feasibility model that quantifies co-inclusion constraints under private mempools.
Second, we examine execution constraints in the absence of builder markets: without guaranteed atomic inclusion, attackers must rely on sequencer ordering, redundant submissions, and priority fee placement, which renders sandwiching probabilistic rather than deterministic.
Third, using transaction-level data from major rollups, we show that naive heuristics overstate sandwich activity.
We find that the majority of flagged patterns are false positives and that the median net return for these attacks is negative.
Our results suggest that sandwiching, while endemic and profitable on Ethereum L1, is rare, unprofitable, and largely absent in rollups with private mempools. 
These findings challenge prevailing assumptions, refine measurement of MEV in L2s, and inform the design of sequencing policies.

%Approximately 95\% of flagged patterns are false positives, as attacker trades do not balance in size, correlations with victim volume are weak, and front--back alignment is inconsistent. 
%Profitability analysis demonstrates that median net returns are negative across all L2s, with typical losses of 0.4--2.2 USD per attempt. Bot-level efficiency remains below 0.05 sandwiches per 100 transactions, indicating that addresses identified as attackers primarily execute arbitrage or other strategies rather than targeted victim exploitation.

%Our results establish that sandwiching, while endemic and profitable on Ethereum L1, is rare, unprofitable, and largely absent in rollups with private mempools. These findings challenge prevailing assumptions, refine measurement of MEV in L2s, and inform the design of sequencing policies.

%% file: sections/01Introduction.tex
\section{Introduction}
    \label{sec:intro}

In March 2025, in block \href{https://etherscan.io/txs?block=22029771}{\texttt{22029771}}, a trader attempted to swap \num{220000}~USDC into USDT on Uniswap v3 on Ethereum. The outcome was unexpected: the trader received only about \num{5000}~USDT. This unfavorable result was due to a sandwich attack.  

Sandwich attacks are among the most detrimental forms of Maximal Extractable Value (MEV)~\cite{daian2020flashboys}, a form of influencing transaction ordering within a blockchain block. In such an attack, an adversary---typically an automated bot---detects a large pending transaction in the mempool and strategically places one transaction immediately before and another immediately after the victim’s. By exploiting predictable price movements, the attacker captures profit while the victim suffers from significantly worse execution. Sandwiching illustrates the harmful side of MEV, though not all MEV is necessarily detrimental~\cite{Heimbach2022MEV}. 

%On Ethereum after the adoption of Proposer–Builder Separation (PBS)~\cite{heimbach2023PBS}, MEV-Boost introduces a regulated market in which block builders compete to maximize block value, and validators are compensated through bids. In the above case, the MEV service was required to advance \num{220000}~USDC as collateral to the builder, leaving only about \num{8000}~USDT in net profit. Validators often highlight such mechanisms as consistent with the ethos of openness and transparency underlying public blockchains.  

Rollups~\cite{sguanci2021layer2,Thibault2022rollups} are popular Ethereum Layer-2 (L2) solutions that primarily aim to improve scalability. However, they may also deploy mechanisms to mitigate (harmful) MEV~\cite{Alipanahloo_2024}.
%Rollups~\cite{sguanci2021layer2,Thibault2022rollups}, Ethereum Layer-2 (L2) solutions, in addition to improving scalability, may also deploy mechanisms to mitigate (harmful) MEV~\cite{Alipanahloo_2024}.
%Rollups~\cite{sguanci2021layer2,Thibault2022rollups} are the leading approaches of Ethereum Layer~2 (L2) scaling.  in addition to improving scalability, they may also deploy mechanisms to mitigate (harmful) MEV~\cite{Alipanahloo_2024}. %% JG: new to address concern.
A common countermeasure against sandwich attacks is the removal of the public mempool in favor of a private pool managed by a centralized sequencer~\cite{chaliasos2024formalrollups}. This approach, adopted by several major L2, prevents adversaries from observing transactions in advance. Nevertheless, early empirical evidence indicates that sandwiching patterns also occur on these L2s, despite private mempools~\cite{torres2024rolling}.

On rollups with fast pre-confirmations (e.g., inclusion receipts), searchers cannot guarantee atomic bundling of front- and back-run transactions, as most sequencers operate private mempools without open builder markets. 
Instead, bots maximize the probability of same-block inclusion via (i) \emph{timing} (predicting proposer cadence and batching windows), (ii) \emph{fee placement} (base fee + priority tips calibrated to sequencer policy), and (iii) \emph{redundancy} (parallelized, nonce-staggered submissions with protective slippage bounds). The absence of public bundle markets shifts MEV from strictly atomic execution to \emph{probabilistic} execution, which we capture in the reward design (Section~\ref{sec:model}).

%\paragraph{Related Work.}
%Qin at al~\cite{qin2021quantifyingblockchainextractablevalue} quantified MEV and sandwich attacks on Ethereum. He analyzed the swap (trade) transaction within the block and identified attacker transactions and match attackers transaction by senders (for user addresses) or recipients (for smart contract). We follow this methodology in our work.

%Other studies\cite{x,y,z} analuzed the impact of snadwich attacks on validatos revenue, and how it depends on size of swaps, typically for Constant Proudct AMM (uniswa v2), We build a model, both for Unsiwa v2 and Concentraitein Liqdy AMM CLL Uniswpa v2), whcih is domineat model on Ehtem and l2s toeay. We then take the perspecigve of an attacker mqimxng his revnue, and show that the optimal straiaon is actually to push the AMM outise of the ticks wiyh lqidytya, which in pracistnce lice in teh examle ealre, is achivned with flash loans.

%\paragraph{Related Work.}

\vspace{0.3em}
\noindent
\textit{Related Work.}
Qin et al.~\cite{qin2021quantifyingblockchainextractablevalue} provide one of the earliest systematic measurements of MEV on Ethereum, including sandwich attacks. 
It identifies attacker trades by examining in-block swap transactions and linking them either through sender addresses (for EOAs) or recipient addresses (for contracts). Our empirical approach follows this identification strategy and adapts it to L2-specific AMM dynamics (such as cyclic arbitrage trades) and shared routers (e.g. Uniswap v4).
Complementary research~\cite{zhou2020high,wang2023collusive} models sandwich profitability under Constant Product AMMs (CPMM) such as Uniswap v2. In contrast, we develop an attacker-profitability model for both CPMMs and concentrated-liquidity AMMs (CLMMs) such as Uniswap v3 and derive an optimal attacker strategy.

The study of MEV on rollups remains at an early stage. Torres et al.~\cite{torres2024rolling} provide the first estimations on L2s, with private-mempool. 
They identified cyclic arbitrage and searched for sandwich attack patterns. 
Their analysis is based on the ERC20 token transfers matching and matching attacker transactions by corresponding sender and receiver.
Due to the mechanics of MEV bot smart contracts, as well as router mechanism introduced by Uniswap (v4), we follow the methodology of Qin et al.~\cite{zhou2020high,wang2023collusive}. for matching attacker transaction and base it on swap events, rather than ERC20 transfers.

Further analysis of MEV on L2 include atomic and non-atomic arbitrage. 
Solmaz et al.~\cite{solmaz2025optimistic} examined cyclic arbitrage execution on L2s by analyzing on-chain bots, while Gogol et al.~\cite{gogol2024crossMEV} and \"{O}z et al.~\cite{oz2025crossMEV} explored non-atomic MEV across rollups. 
\"{O}z et al.~\cite{Oz2024MEV} highlights that on First Come, First Served (FCFS) L1s transaction-ordering techniques commonly exploited for MEV extraction on blockchains with fee-based prioritization do not directly translate to systems where ordering is determined solely by transaction arrival times.

%\todo[inline]{JG: cite~\cite{DBLP:conf/sp/QinZG22} for reviewer 50A. [DONE]}
%\todo[inline]{JG: 50B: how do our \emph{analyses} differ from each of these? [DONE]}

\vspace{0.3em}
\noindent
\textit{Contributions.}
%\paragraph{Contributions.}
This paper advances the understanding of MEV in rollups with private mempools. Our main contributions are:

\textbf{(i) Empirical.}
We conduct an empirical study of sandwich activity across major rollups with private mempools, using strict attacker-linking heuristics that distinguish EOAs, routers, and smart accounts~\cite{qin2021quantifyingblockchainextractablevalue}. 
To our knowledge, this is the first measurement of L2 sandwich activity that applies this methodology and \emph{swap-event} analysis rather than ERC20 token-transfer inference. 

Across all L2s, we find no evidence of sustained or economically meaningful sandwich attacks.  
Naive pattern-based heuristics vastly overstate activity: over 95\% of flagged triples fail economic consistency checks, median net profit-and-loss (PnL) is negative, and sandwich efficiency remains below 0.05 per 100 transactions.

 \textbf{(ii) Theoretical.}
    We extend the classical CPMM model to concentrated-liquidity AMMs. We show that the attacker optimal strategy depends on the tick boundaries and the liquidity distribution. In particular, attackers benefit most when pushing trades into thinner liquidity regions, which is in practice executed with flash-loans. We complement this framework with an execution-feasibility model that quantifies co-inclusion constraints under private mempools.

 \textbf{(iii) Execution feasibility under private mempools.}
    We analyze the execution constraints imposed by centralized sequencers and private mempools. By formalizing co-inclusion probabilities as a function of block time and tip, we show that L2 sandwiching is inherently probabilistic rather than atomic, sharply limiting practical feasibility in today’s rollup architectures.

\textit{Taken together, our findings inform rollup and sequencer design: the empirical absence of sandwiches on current L2s arises from private mempools and medium-value swaps rather than from inherent structural protections. As ecosystems transition toward public mempools, order-flow auctions, or higher-value trades, sandwich attacks may re-emerge, and sequencing policies should be evaluated with these economic risks explicitly in scope.}

%\subsubsection{Paper Organization.} 
%Section~\ref{sec:background} summarizes L2 sequencing and related work.
%Section~\ref{sec:model} formalizes the agent and reward functions.
%Section~\ref{sec:setup} details the experimental setup.
%Section~\ref{sec:results} presents results.
%Section~\ref{sec:discussion} discusses implications.
%Section~\ref{sec:limitations} covers limitations and future work.
%Section~\ref{sec:conclusion} concludes.

%% file: sections/03Model_v3.tex
\section{Sandwich Profitability and Execution Probability}
\label{sec:model}

Sandwich attacks exploit predictable price impact: an adversary buys before a victim's X$\rightarrow$Y swap and sells after the victim moves the price upward. This section provides a concise analytical model for optimal attack sizing under
both constant-product (CPMM) and concentrated-liquidity (CLMM) AMMs, followed by a quantitative execution-feasibility model for private-mempool rollups. %Detailed derivations and proofs are provided in Appendix~\ref{app:proofs}.

\subsection{Economic Model}
\label{sec:eco-model}

We model sandwich profitability across CPMMs and CLMMs in the small-trade
regime characteristic of today’s rollups, where both attacker and victim trades
are typically small relative to tick-level liquidity. The objective is to derive
the attacker’s optimal frontrun size $V_f$ as a function of the victim's input
$V_v$, the AMM's liquidity profile, and the victim's slippage constraint.
Throughout, we measure profits in USD-equivalent units. 
%Proofs of all results are provided in Appendix~\ref{app:proofs}.

\paragraph{Setting.}
An attacker executes a frontrun trade (X$\to$Y), followed by the victim’s
X$\to$Y trade, and completes with a backrun (Y$\to$X). Let $\phi$ denote the
proportional swap fee and $L$ the effective liquidity depth of the AMM at the
current price. In CPMMs, $L$ corresponds to the reserve $x_0$; in CLMMs, $L$
corresponds to the tick-level liquidity $L_i/P$. In both cases, the attacker’s
profit arises from the temporary wedge in exchange rate generated by the
victim’s price impact.

\paragraph{Small-trade regime.}
Define the normalized input sizes
\begin{equation}
\alpha_f = \frac{(1-\phi)V_f}{L},
\qquad
\alpha_v = \frac{(1-\phi)V_v}{L},
\end{equation}
and assume $\alpha_f,\,\alpha_v \ll 1$, which empirically holds for most
candidate patterns on rollups. Expanding the AMM swap function to second
order in $(\alpha_f,\alpha_v)$ yields the following result.

\begin{proposition}[Small-trade approximation on CPMMs]
\label{prop:cpmm}
Consider\\
a constant-product AMM with proportional fee $\phi$ and depth $L$. 
Let $V_f$ and $V_v$ denote the attacker and victim inputs. In the small-trade 
regime, the attacker's incremental profit (conditioning on the presence of the 
victim) admits the quadratic approximation
\begin{equation}
\label{eq:quad-profit}
\Delta \Pi(V_f;V_v)
\;\approx\;
\frac{(1-\phi)^2}{L} \big( V_f V_v - V_f^2 \big)
\;-\; 2\phi V_f,
\end{equation}
where the term $2\phi V_f$ represents the swap fees paid on the frontrun and backrun legs. The unique interior maximizer is
\begin{equation}
\label{eq:halfrule}
V_f^\star = \frac{V_v}{2}.
\end{equation}
The effective optimizer is 
$V_f^\star = \min\{\,V_v/2,\;V_f^{\max}\,\}$,
where $V_f^{\max}$ is the upper bound implied by the victim's slippage tolerance.
\end{proposition}

Proposition~\ref{prop:cpmm} recovers the classical ``half-the-victim'' rule: the frontrun should be approximately half as large as the victim trade.
%This result provides the benchmark structure against which empirical scaling relationships can be compared. 
As shown later, most L2 candidate sandwiches  fail this proportionality condition, suggesting non-adversarial origins.

\paragraph{Extension to CLMMs.}
Concentrated-liquidity AMMs partition the price axis into ticks 
$[P_i,P_{i+1}]$, each endowed with constant liquidity $L_i$. Inside a single 
tick, the local depth is $L=L_i/P$, and the CPMM approximation applies 
verbatim. However, unlike CPMMs, CLMM liquidity is piecewise constant across 
ticks, producing discontinuities in marginal price impact at tick boundaries.

\begin{proposition}[Optimal frontrunning across CLMM ticks]
\label{prop:clmm}
Let the current price lie in tick $[P_i,P_{i+1}]$ with liquidity $L_i$.  
If the combined effective input of the frontrun and victim remains inside this 
tick, the optimal frontrun is the CPMM solution $V_f^\star = \min\{V_v/2,\,
V_f^{\max}\}$.  
If the combined path reaches the boundary and the next tick has lower liquidity 
($L_{i+1} < L_i$), the attacker strictly benefits from nudging the victim into 
the thinner region. In that case, the global optimum is
\begin{equation}
V_f^\star = \min\{ V_f^{\mathrm{gap}},\;V_f^{\max} \},
\end{equation}
where $V_f^{\mathrm{gap}}$ is the minimal frontrun size required to push the 
joint flow across the boundary.  
If $L_{i+1} \ge L_i$, no profit jump occurs and the within-tick optimizer 
remains optimal.
\end{proposition}

Proposition~\ref{prop:clmm} shows that CLMM sandwiches are
\emph{piecewise quadratic}: CPMM-like inside ticks, but with discrete 
profit jumps at boundaries with declining liquidity.  
This structure implies that genuine CLMM sandwiches should exhibit highly
specific scaling patterns—either near $V_f\approx V_v/2$ (in-tick) or clustering
around $V_f^{\mathrm{gap}}$ (boundary crossing). Deviations from both patterns,
as shown empirically in Section~\ref{sec:empirical}, are indicative of false
positives or non-sandwich MEV flows.

\paragraph{Empirical profit approximation.}
The small-trade CLMM profit before gas fees can be expressed compactly as
\begin{equation}
\Pi_{\text{net}} \;\approx\; (1-\phi)^{2}\,\frac{\varepsilon}{L}\,\big(V_f V_v - V_f^{2}\big)\;-\;2V_f\phi,
\label{eq:empirical-pnl}
\end{equation}
%\todo[inline]{50b: does this incldue tx fees? KG: No, this formula is for AMM slippage/price impact, I subtract fees separately further in this section. I will clarify this.}
\noindent
where $V_v$ denotes the victim’s input, $V_f$ the attacker’s input per leg, $\phi$ the proportional swap fee (e.g.\ $\phi=0.0005$ for 5~bps), $L$ the real liquidity available in the active tick, and $\varepsilon \approx 1.0001^{\Delta \mathrm{tick}} - 1$ the relative width of the tick. 
For the common case of $\Delta \mathrm{tick}=1$ and small $\phi$, this reduces to
\[
\Pi_{\text{net}} \;\approx\; \frac{V_f(V_v-V_f)}{L}\cdot 10^{-4}\;-\;2V_f\phi,
\]
which we use as the empirical profit approximation before gas fees to assess the (un)profitability of candidate sandwiches in observed data. Further, we report profits both before after gas fees.

% ----------------------------------------------------------------------

\subsection{Execution Probability Under Private Mempools}

Even if the economic optimum $V_f^\star$ is well defined, executing a sandwich requires that the attacker’s frontrun ($f$), victim transaction ($v$), and backrun ($b$) all appear in the same block (or adjacent micro-batches) and in the correct order. On rollups with private mempools, no builder market guarantees atomic inclusion of $(f,v,b)$. Instead, execution becomes
\emph{probabilistic} and depends critically on the sequencer’s ordering rule. Most L2s implement two canonical policies:

\begin{itemize}
    \item \textit{(1) First Come, First Served (FCFS; time-priority):} transactions are ordered primarily by arrival time, with tips used only as tie-breakers in micro-batches,
    \item \textit{(2) Priority Gas Acution (PGA; tip-priority):} transactions within a batch are ordered by descending priority fee. Arrival time affects only which batch a transaction enters.
\end{itemize}

This distinction is central: under FCFS, attackers optimize \emph{timing} ($\Delta$), while under PGA they optimize \emph{tips} ($t_f,t_b$) and face a Bayesian ``tip-guessing'' problem relative to the victim and background flow.
We now formalize the components of execution probability in a framework that accommodates both rules. Let
\begin{itemize}
    \item $T_b$ = block time,
    \item $T_s$ = sequencer batching window,
    \item $\Delta$ = time between $f$ and $b$ submissions (attacker choice under FCFS),
    \item $t_f, t_v, t_b$ = priority tips on front-, victim-, and back-run,
    \item $F(\cdot)$ = sequencer ordering function ($F(t)$ increasing in tips under PGA),
    \item $\sigma$ = network/latency variance.
\end{itemize}

% ---------------------------------------------------------------------
\vspace{0.5em}
\noindent
\textbf{Same-batch probability.}
For both FCFS and PGA, $f$ and $b$ must fall into the same batch for a single-block sandwich. A simple approximation is
\begin{equation}
p(\mathrm{batch})
\;\approx\;
\max\!\left(0,\,1-\frac{\Delta}{T_s}\right).
\end{equation}
Under FCFS, $\Delta$ is a strategic choice; under PGA, attackers typically send
$f$ and $b$ simultaneously, so $\Delta \approx 0$.

% ---------------------------------------------------------------------
\vspace{0.5em}
\noindent
\textbf{Priority-ordering probability: FCFS vs.\ PGA.}
A valid sandwich requires $f \prec v \prec b$ under the sequencer’s internal ordering rule. Even conditional on $t_f > t_v > t_b$, the ordering constraint
$f \prec v \prec b$ may fail if background transactions with intermediate
priority fees are inserted between the attacker’s legs.

\paragraph{FCFS.}
Ordering is determined (up to noise) by arrival times. Let $N$ be the number of intervening transactions between submission times of $f$ and $b$. If background transactions arrive according to a Poisson process with rate $\lambda$, then
\[
p_{\rm FCFS}(\mathrm{priority})
=
\Pr[N=0]
=
e^{-\lambda \Delta}.
\]
Here the attacker optimizes $\Delta$: shorter delays reduce the probability of
interference.

\paragraph{PGA.}
Ordering is determined by priority tips. A correct sandwich requires
$t_f > t_v > t_b.$
Let background tips be i.i.d.\ from distribution $G$. If $K$ transactions enter
the batch with $v$, then the probability that none fall in the dangerous
interval $(t_b,t_f)$ is
\[
p_{\rm PGA}(\mathrm{priority})
=
\mathbb{E}_{t_v,K}\!\big[
\mathbf{1}\{t_f > t_v > t_b\}
(1 - (G(t_f)-G(t_b)))^{K}
\big].
\]
Under PGA, timing does not improve relative ordering; only tip selection matters, and the attacker faces a Bayesian optimisation over $(t_f,t_b)$ given uncertainty about $t_v$.

% ---------------------------------------------------------------------
\vspace{0.5em}
\noindent
\textbf{Arrival-time probability.}
Arrival times are noisy. If the effective delay between intended and realized arrival times is Gaussian with variance $\sigma^2$, then
\begin{equation}
p(\mathrm{arrival})
\;\approx\;
\exp\!\left(-\frac{\Delta^2}{2\sigma^2}\right).
\end{equation}
This component is relevant primarily under FCFS, where $\Delta$ is a choice variable; under PGA, $\Delta$ is typically negligible.

% ---------------------------------------------------------------------
\vspace{0.5em}
\noindent
\textbf{Co-inclusion probability.}
Combining the components yields the attacker's feasibility frontier:
\begin{equation}
\label{eq:coinc}
p(\mathrm{co\text{-}inc})
\;\approx\;
\left(1-\frac{\Delta}{T_s}\right)
p_{\rm policy}(\mathrm{priority})
\exp\!\left(-\frac{\Delta^2}{2\sigma^2}\right),
\end{equation}
where $p_{\rm policy}$ is $p_{\rm FCFS}$ or $p_{\rm PGA}$ depending on the
sequencer rule.

For typical rollup parameters ($T_b \approx 200$--500\,ms, $T_s \approx 300$--800\,ms, $\sigma \approx 50$\,ms), the resulting co-inclusion probabilities lie between $0.05$ and $0.20$ across both FCFS and PGA systems, well below the near-atomic inclusion achievable on Ethereum L1 via bundle markets. Consequently, even economically optimal sandwiches have low expected value on today’s private-mempool L2 architectures.

%\paragraph{Takeaway.}
%The economic model identifies an optimal frontrun $V_f^\star = V_v/2$ (or a boundary-nudging solution in CLMMs), but the execution-feasibility model shows that obtaining the required ordering is intrinsically probabilistic on rollups without builder markets. This tension explains the empirical absence of profitable sandwiches despite  their theoretical existence.

\subsection{Minimum Victim Swap Size for Positive Expected Value on L2s}
\label{sec:minsize}

The attacker’s expected profit on a rollup with private mempools is the product of (i) the economic profit generated by the AMM price impact and (ii) the probability of co-inclusion under the sequencer’s ordering rule. From Section~2.1, under the small-trade approximation the attacker’s optimal frontrun size satisfies $V_f^\star \approx V_v/2$, yielding gross profit
\begin{equation}
\Pi_{\mathrm{gross}}^\star(V_v)
\;\approx\;
\frac{1}{4L} V_v^{2},
\end{equation}
where $L$ is the effective liquidity depth of the active tick in the CLMM. However, as shown in Section~2.2, in private-mempool rollups the sandwich executes only with probability $p_{\mathrm{succ}} \in [0.05, 0.20]$, depending on batching windows, tip ordering, and arrival-time variance. The attacker’s expected value (EV) satisfies
\begin{equation}
\mathbb{E}[\Pi]
=
p_{\mathrm{succ}}
\cdot
\Pi_{\mathrm{gross}}^\star(V_v)
-
C_{\mathrm{gas}} - C_{\mathrm{slip}},
\end{equation}
where $C_{\mathrm{gas}}$ denotes L2 gas fees for the front- and back-run (typically
\$0.20--\$0.70 in total) and $C_{\mathrm{slip}}$ is the slippage loss incurred by
the attacker’s own trades.

A sandwich becomes profitable only when $\mathbb{E}[\Pi] > 0$, which implies the following minimum victim trade size:
\begin{equation}
V_v^{\min}
\;=\;
2 \sqrt{ L \cdot \frac{C_{\mathrm{gas}} + C_{\mathrm{slip}}}{p_{\mathrm{succ}}} }.
\label{eq:minimumV}
\end{equation}

Under co-inclusion probabilities $p_{\mathrm{succ}} \in [0.05, 0.20]$ and typical active tick liquidity $L$ between \$50k and \$300k, Eq.~\eqref{eq:minimumV} implies that victim swap sizes must generally exceed \$1.5k--\$3k.
These thresholds exceed by a factor of $2\times$--$10\times$ the median victim swap sizes observed in Section~5 for major rollups (typically \$200--\$1{,}200). Consequently, even economically optimal sandwiches would yield negative expected value on today’s L2s. %gross gains scale quadratically with $V_v$, whereas (i) co-inclusion probability remains low, (ii) gas fees are non-negligible, and (iii) CLMM liquidity is shallow. This theoretical barrier aligns with the empirical results of Section~5, where over 95\% of candidate patterns fail economic consistency and median net PnL is negative across all chains.

%% file: sections/04Enforcing-order.tex
\subsection{Execution Feasibility}
%\todo[inline]{JG: I will edit to address concerns of reviewer 50A.}

Executing a sandwich attack requires that a transaction $f$ front-runs a set of victim transactions $v_i$ both of which are followed by a back-running transaction $b$.
Here, we outline methods by which someone can attempt a sandwich attack.

While it is relatively easy to establish that $f$ should be the first transaction of a block (e.g., by paying the highest priority fee), it is more difficult to ensure that $b$ comes after some victim transactions.
Necessarily, the attacker must send $f$ and $b$ in two different transactions to the target blockchain and the nonce of $b$ must be higher than the nonce of $f$ if $b$ is to be considered valid.
This is also sufficient though it does not guarantee that both $f$ and $b$ appear in the same block.
Assuming that block builders sort by priority fee, it is necessary to have that the priority fee of $b$ is not larger than the priority fee of $f$ (and larger fee gaps provide more opportunities for victim transactions between $b$ and $f$).
In order to execute $b$ after $f$, two straightforward approaches could be used:

\begin{itemize}
    \item Introducing an artificial delay before sending $b$ to the blockchain. This may work under FCFS-style sequencing on blockchains with relatively slow block production times (e.g., block times greater than $1$s), but is generally ineffective under PGA-style ordering, where delaying $b$ weakens batch co-inclusion without improving relative ordering. On fast-rollup chains (e.g., Base with $\sim200$ms block times), such delays often prevent same-block execution altogether.
    \item Sequentially submitted calls without a delay but different tip values via an intermediary contract $C$ which contains the necessary logic for both front-running and back-running.
    A function may be placed behind a modifier which chooses which logic to execute based on the contract state and current block.
    %The modifier may increment a counter with after every call, choosing to execute a front-running transaction if the counter is even when the function is called or executing a back-running transaction if the counter is odd.

        %%% jgorzny: No! Sadly, the following approach _only_ works if the logic of f and b are executed in the same transaction. That would would mean that all victim transactions are also exected by the attacker. That seems unlikely. There is at least one corner case: if you are calling someone's smart wallet (via AA primitives), you could use this to sandwich their trades. But I guess they wouldn't trust you to do that for much longer afterwards...
        %\item The modifier may make use of Cancun's transient memory to ensure that back-running logic is only executed after front-running logic.
        %This approach can be simulated without Cancun's opcodes using methods described in~\cite{DBLP:conf/dappcon/CallensMG24}.
\end{itemize}

\begin{comment}

A typical statistical sandwich bot proceeds in two waves:
\begin{enumerate}
  \item \textbf{Front-run wave:} Submit a set of split and duplicated
  transactions of size $q_t^f$ into the pool before the predicted victim.
  These are parameterized by a splitting factor $n_t^f$ and a duplication
  factor $K_t^f$, generating $n_t^f K_t^f$ attempts. Each attempt is wrapped
  with a tight block guard 
  ($\text{minBlock}=\text{current}+1$, $\text{maxBlock}=\text{current}+1$) 
  and a short deadline (e.g., 3--5 seconds), so that if it misses the intended
  block it reverts at low cost. Priority tips $\tau_t^f$ are chosen from a grid
  biased toward early slots (e.g., Slot~1 in Flashblocks, express lane in 
  TimeBoost).
  
  \item \textbf{Back-run wave:} After a short delay (50--150 ms), the bot 
  submits the back-run legs of size $q_t^b$, split into $n_t^b$ slices and 
  duplicated $K_t^b$ times. The back-run includes price guards, ensuring that 
  execution occurs only if the victim trade has moved the pool into the 
  expected post-trade price band. Tips $\tau_t^b$ are biased toward later slots 
  (e.g., Slot~8 in Flashblocks, standard lane under TimeBoost), so that the 
  back-run legs are more likely to settle after the victim.
\end{enumerate}

\end{comment}

%% file: sections/05Methodology.tex
\section{Methodology}
\label{sec:methodology}

\textbf{Data Collection.}
We collected all swap event logs with corresponding transactions and transaction traces from Arbitrum, Base, Optimism, Unichain and ZKsync from January to September 2025, using Dune queries. %. We used Dune queries for data collection.
For profit analysis, we extract pool-level liquidity (for CPMMs), current and surrounding ticks liquidity (for CLMMs), and the applicable LP fee tier for each pool.

\vspace{0.3em}
\noindent
\textbf{Sandwich Pattern Heuristics.}
A central challenge in detecting sandwich attacks is reliably linking the front- and back-running legs to the same adversarial actor. A naïve method is to use the \texttt{taker} field from pool-level \texttt{Swap} events, which records the immediate caller of the pool contract. While convenient, this has two major drawbacks: (i) it often resolves to shared routers or aggregators (e.g., UniswapRouter, 1inch) that serve many independent users, and (ii) it may differ across the front- and back-leg if the attacker routes through multiple entrypoints. As a result, \texttt{taker}-based attribution misses valid sandwiches and produces false positives. For instance, when shared routers collapse unrelated users into a single calling address, two benign trades may appear as a coordinated front–back pair. 
%\todo[inline]{50B: Why false positives? (answered above)}
To overcome this, we derive an adversary identifier $\mathsf{actor\_id}$ (as described in Step 3 below) and  detect candidate sandwich patterns using the following procedure: %from transaction-level fields:

\begin{enumerate}
  \item \textbf{Pre-filtering of atomic arbitrage.} Exclude single transactions with cyclic (atomic) arbitrage transactions, as they may mimic sandwich legs. We follow the methodology of~\cite{torres2024rolling} for detection.
  %\todo[inline]{50B: how were these identified? JG: do you just mean that single tx arbitrate was ignored? If so, I can clarify that.}

\item \textbf{Partitioning.} Group all swaps by $(\texttt{blockchain}, \texttt{block\_number}, \mathsf{pool\_address})$ so that only transactions interacting with the same liquidity pool in the same block are compared. For Uniswap v4, we extend the grouping with the currency pair, since event logs register only the router address.

\item \textbf{Actor attribution.} %Assign each transaction an $\mathsf{actor\_id}$ according to the above rule, ensuring that front- and back-run legs routed through different routers are still linked, while excluding the victim’s own address.
Assign each transaction an $\mathsf{actor\_id}$ according to the following rule.
If the top-level callee \texttt{tx\_to} is a known system contract (e.g., ZKSync bootloader) or a widely used router/aggregator, we set $\mathsf{actor\_id} := \texttt{tx\_from}$, which refers to the funding externally owned account (EOA) or smart account. This captures adversaries who execute via shared infrastructure.
Otherwise, if the callee \texttt{tx\_to} is a custom contract, we set $\mathsf{actor\_id} := \texttt{tx\_to}$. This captures adversaries using dedicated executor contracts.

\item \textbf{Ordering constraint.} Within each group, order transactions by block index and select triples $(f, v_i, b)$ that satisfy $\text{index}(f) < \text{index}(v_i) < \text{index}(b)$ and $\mathsf{actor\_id}(f) = \mathsf{actor\_id}(b) \neq \mathsf{actor\_id}(v_i)$.

%\[
%\text{index}(f) < \text{index}(v_i) < \text{index}(b),
%\quad \mathsf{actor\_id}(f) = \mathsf{actor\_id}(b) \neq \mathsf{actor\_id}(v_i).
%\]

\item \textbf{Swap-direction constraint.} Require that the trade direction of the victim and front legs match, while the back leg has the reverse direction.

\end{enumerate}

%This step-wise heuristic improves recall compared to the naïve \texttt{taker}-based method while maintaining high precision. It also generalises across rollups: in account-abstraction systems such as zkSync, nearly all \texttt{tx\_to} values are system contracts, so $\mathsf{actor\_id}$ defaults to the funding account, whereas in Arbitrum or Optimism many attackers deploy bespoke executors, for which \texttt{tx\_to} is already stable.

%% file: sections/06EmpiricalResults.tex
\section{Empirical Evidence}
\label{sec:empirical}

%We begin with a descriptive overview of the candidate sandwich triples identified across all swap events on studied rollups. This section characterizes transaction sizes, bot activity, and attack profit and loss (PnL).

%This section characterizes candidate sandwich triples, bot activity, and attack profit and loss (PnL) identified across all swap events on studied rollups.

Before presenting the empirical evidence, we briefly connect the theoretical results of Section~\ref{sec:model} to the measurement strategy. The quadratic profit model predicts that genuine sandwiches exhibit two key signatures: (i) attacker legs scale proportionally with the victim ($V_f^\star \approx V_v/2$ inside a tick), and (ii) front- and back-run sizes should be closely matched except at tick boundaries with declining liquidity.  
At the same time, the execution-feasibility model implies that on private-mempool rollups only a small fraction of attempted sandwiches can achieve correct ordering, even when economically optimal.  

These theoretical constraints motivate the economic-consistency checks used in this section and explain why most observed ``sandwich-shaped'' triples are expected to be false positives or arbitrage by-products rather than profit-maximizing attacks.

%\todo[inline]{50B: doe we consider other pools? no -- we should say something about why not. KG: We analyze all pools that are labelled by Dune as AMM pool, so in practice almost everything. I will clarify.}

\subsection{Swap Size Analysis}

Table~\ref{tab:descriptive-sizes-split} reports descriptive statistics of swap sizes for candidate sandwich triples across studied rollups. Base dominates the sample with over 50{,}000 identified triples, while Optimism and ZKsync have limited coverage. 

\begin{table}[tb]
\centering
\sisetup{
  round-mode=places,
  round-precision=1,
  scientific-notation=false,
  group-separator={,}
}
%\label{tab:descriptive-sizes}
\caption{Summary statistics of swap sizes (USD) for front, victim, and back legs. For each role we report the median and the interquartile range (IQR).}
\label{tab:descriptive-sizes-split}
\begin{tabular}{l r r r r r r r}
  \toprule
  & & \multicolumn{3}{c}{Median} & \multicolumn{3}{c}{IQR} \\
  \cmidrule(lr){3-5} \cmidrule(lr){6-8}
  Chain & Count & Front & Victim & Back & Front & Victim & Back \\
  \midrule
  Arbitrum & 2,576  & 807.1  & 872.8  & 1077.1 & [400.3, 1960.7] & [403.2, 2391.9] & [500.4, 2662.7] \\
  Base     & 50,952 & 2230.5 & 1224.0 & 2157.7 & [911.9, 4060.5] & [398.8, 2999.0] & [905.8, 3825.0] \\
  Optimism & 177    & 464.1  & 1103.0 & 309.2  & [211.4, 1469.8] & [447.9, 3167.1] & [179.5, 640.5] \\
  Unichain & 795    & 682.5  & 3045.4 & 546.8  & [273.1, 3662.3] & [500.7, 6449.8] & [222.9, 2934.0] \\
  ZKsync   & 27     & 441.4  & 400.6  & 194.2  & [255.8, 665.3]  & [249.4, 1729.1] & [153.2, 430.0] \\
  \bottomrule
\end{tabular}
\end{table}

Across chains, the relative sizing of victim and attacker legs is inconsistent. On Base, for instance, the median victim size is 
\$1.2k, while the corresponding front- and back-runs are larger, around \$2.2k and \$2.1k respectively. On Optimism and ZKsync, front- and back-runs are frequently much smaller than victims, while on Unichain the opposite pattern appears, with victims typically several times larger than attacker legs. Importantly, across all chains, these swaps are small in absolute terms,  with median values on the order of only a few hundred to a few thousand USD, far below the scale typically associated with  profitable sandwiches on Ethereum L1~\cite{qin2021quantifyingblockchainextractablevalue,Torres2021Frontrunning}.
This lack of consistent scaling between attacker and victim swap sizes, combined with their generally modest magnitudes, foreshadows our later finding that most of these triples are not genuine sandwiches but are better explained by arbitrage activity or unrelated transactions coinciding in the same block.

%\subsection{False Positives}
\vspace{0.5em}
\noindent
\textbf{False Positives.} 
Table~\ref{tab:strong-sig} presents cross-chain statistics on candidate sandwich triples. ``Strong Sig.'' denotes cases, in which backrun swap size does not deviate by more than $10\%$ from the frontrun swap. Reported correlations are Pearson coefficients. The relatively low share of Strong Signature cases (5--25\%) and consistently weak correlations among front-, back-running and victim swaps indicate that most candidate triples are not genuine victim-targeted sandwiches.

\begin{figure}[p]
\makebox[\textwidth][c]{%
  \includegraphics[width=\linewidth]{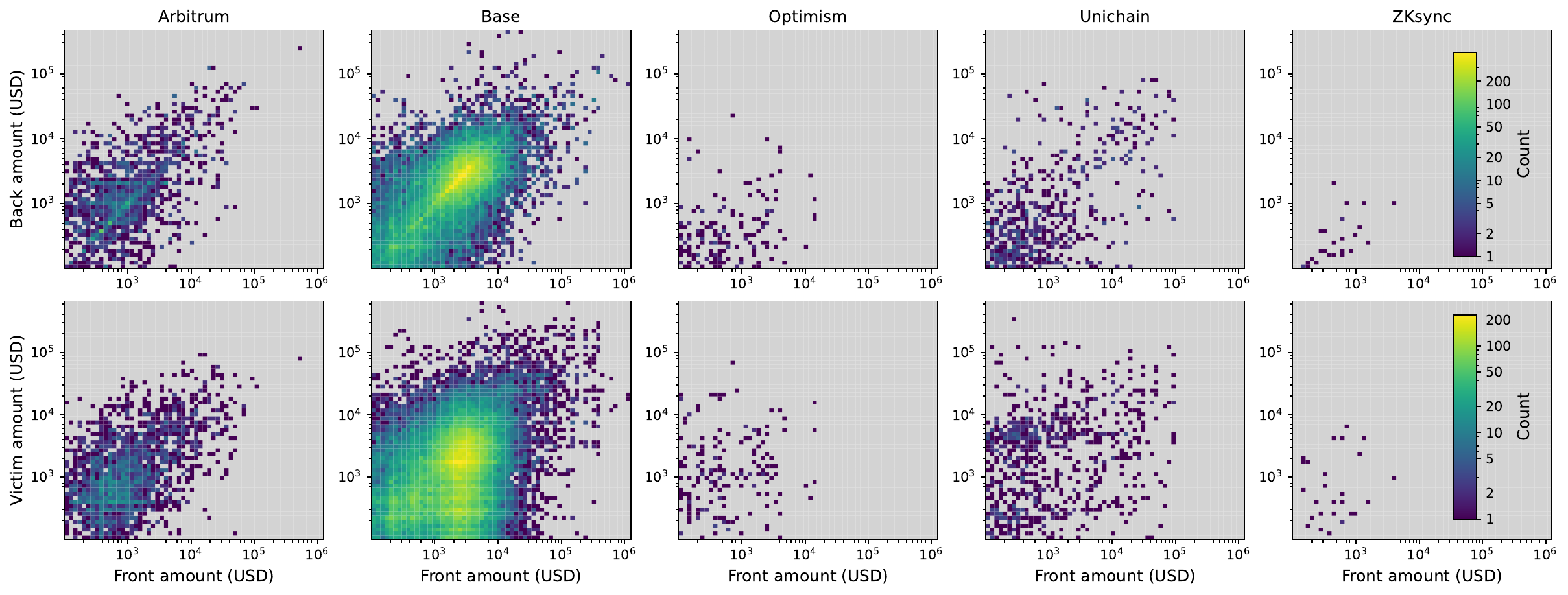}}
\caption{Density heatmaps of swap sizes (USD) for front, victim, and back legs. The top row plots back-run versus front-run swaps; a genuine sandwich would concentrate mass along the diagonal $b=f$, yet observed densities scatter widely away from it. The bottom row plots victim versus front-run swaps; here too, the absence of clustering along the diagonal $v=f$ indicates that attacker trades do not scale with victim size.}
\label{fig:heatmap-back-front}
\end{figure}

\vspace{0.5em}
\noindent
\textbf{Back vs. Front Size Mismatch.} 
Median back/front ratios deviate substantially from $1.0$, with wide dispersion across chains. For example, the median ratio is $0.728$ on Optimism (IQR $[0.282, 1.526]$) and $0.825$ on Unichain (IQR $[0.358, 1.913]$). True sandwiches should cluster near $1.0$ (back $\approx$ front). Instead, attackers leave large mismatches, consistent with external hedging such as a CEX leg. Figure~\ref{fig:heatmap-back-front} highlights these deviations, showing distributions dispersed and biased away from zero.

\vspace{0.5em}
\noindent
\textbf{Front vs. Back Correlation is Weak.} 
Front$\leftrightarrow$back correlations remain far below $1.0$. For instance, the correlation is only $0.08$ on Optimism and $0.40$ on Unichain, with the highest case $0.63$ on Arbitrum. True sandwiches require matched trade sizes, but the observed weak dependence suggests that front and back trades are not coordinated pairs, rather independent legs of broader arbitrage strategies.

\vspace{0.5em}
\noindent
\textbf{Victim Trade has Little Influence on Attacker Size.} 
Correlations between victim trades and attacker trades are consistently weak. Victim$\leftrightarrow$front correlations are as low as $0.03$ on Optimism and only $0.16$ on Unichain; victim$\leftrightarrow$back correlations reach $0.18$ on Unichain and $0.57$ on ZKsync but remain far from 1. In genuine sandwiches, attacker trades should scale with victim size. The absence of such scaling indicates that attacker activity is not driven by exploiting victim slippage.

%\begin{figure}[htbp]
%\centering
%\includegraphics[width=0.7\linewidth]{figures/violin_back_front_deviation.pdf}
%\caption{Distribution of back- front- running swaps size deviations (in \%). A genuine sandwich would cluster around $0\%$, whereas the observed distributions are wide and systematically biased away from zero, suggesting that most candidate triples do not represent victim-targeted sandwiches.}
%\label{fig:violin-back-front}
%\end{figure}

\begin{table}[tbp]
\centering
\small
\caption{Cross-chain summary of candidate sandwich triples. ``Strong Sig.'' denotes cases of back/front (b/f) size matching criteria. Reported correlations are Pearson coefficients. The relatively low shares of Strong Signature cases and weak correlations across chains suggest that most candidate triples are not genuine victim-targeted sandwiches.}
\label{tab:strong-sig}
\setlength{\tabcolsep}{6pt}
\renewcommand{\arraystretch}{1.2}
\begin{tabular}{l r r l r r r}
\toprule
Chain & Strong Sig. & Median b/f & IQR b/f & 
\makecell{Victim\\$\leftrightarrow$Front} & 
\makecell{Victim\\$\leftrightarrow$Back} & 
\makecell{Front\\$\leftrightarrow$Back} \\
\midrule
Arbitrum  & 24.6\% & 1.035 & [0.816, 2.672] & 0.426 & 0.409 & 0.627 \\
Base      & 14.3\% & 0.967 & [0.549, 1.583] & 0.276 & 0.375 & 0.352 \\
Optimism  & 5.1\%  & 0.728 & [0.282, 1.526] & 0.032 & 0.592 & 0.082 \\
Unichain  & 7.3\%  & 0.825 & [0.358, 1.913] & 0.162 & 0.182 & 0.398 \\
ZKsync    & 7.4\%  & 0.674 & [0.392, 0.785] & 0.081 & 0.567 & 0.354 \\
\bottomrule
\end{tabular}
\label{tab:sandwich-summary}
\end{table}
\begin{figure}[tp]
  \centering
  \includegraphics[width=\linewidth]{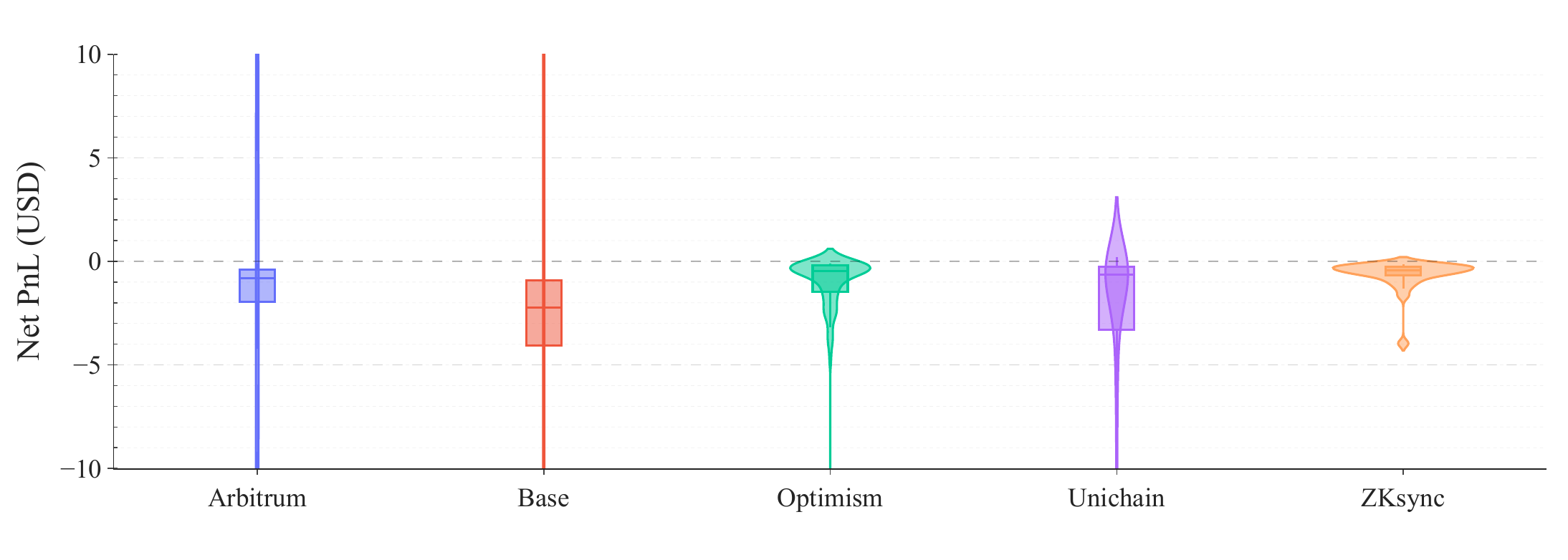}
 % \fbox{\rule[0pt]{0pt}{2in} \hspace{3in}} % placeholder box
  \caption{Violin distribution of estimated net PnL in USD. The distributions are tightly centered around zero, with heavy tails, confirming that median profits are negligible and most candidate sandwiches are not profitable.}
  \label{fig:violin-pnl}
\end{figure}

\subsection{Negative Net Profits}
We report three profitability measures for each candidate sandwich attack. The first, \emph{median\_pnl\_gross\_usd}, represents the potential gain from exploiting the victim’s price impact before accounting for the attacker’s own costs. The second, \emph{median\_pnl\_net\_slippage\_usd}, adjusts this value by incorporating the attacker’s own slippage losses, capturing the fact that executing front- and back-run trades also moves the price against them. The third, \emph{median\_pnl\_net\_usd}, further deducts transaction fees and observed gas expenditures, providing the most conservative estimate of realized profitability.

Empirically, as summarized in Table~\ref{tab:pnl-summary}, median gross profits are on the order of $10^{-3}$ USD per attack and are almost entirely eliminated once slippage is considered, with final median net PnL turning negative across all rollups. Figure~\ref{fig:violin-pnl} further illustrates that these values are tightly clustered around zero with heavy tails, reinforcing that candidate sandwiches do not generate sustainable profits in practice and that the vast majority of observed patterns are better explained by arbitrage-like or other MEV activity rather than deliberate victim targeting.

\begin{table}[p]
\centering
\sisetup{
  round-mode=places,
  round-precision=4,
  scientific-notation=false,
  group-separator={,}
}
\setlength{\tabcolsep}{10pt} % default ~6pt

\caption{Summary of estimated Profit-and-Loss (PnL) in USD per candidate sandwich across rollups. Gross PnL is the attacker’s profit before costs, Net Slippage PnL accounts for execution price impact only, and Net PnL further deducts gas fees. Medians across all L2s are close to zero, and interquartile ranges (IQR) for Net PnL confirm that most sandwich patterns cluster around small losses.}

\label{tab:pnl-summary}
\begin{tabular}{l r r r l}
  \toprule
  Chain & \makecell{Median\\Gross PnL} & \makecell{Median\\Net Slippage} & \makecell{Median\\Net PnL} & \makecell{IQR\\Net PnL} \\
  \midrule
  Arbitrum  & \num{0.000480} & \num{0.000010}  & \num{-0.802645} & [\num{-1.962861}--\num{-0.400186}] \\
  Base      & \num{0.001811} & \num{-0.000535} & \num{-2.227808} & [\num{-4.059283}--\num{-0.910586}] \\
  Optimism  & \num{0.001017} & \num{0.000140}  & \num{-0.456947} & [\num{-1.466618}--\num{-0.206900}] \\
  Unichain  & \num{0.002731} & \num{0.000350}  & \num{-0.652108} & [\num{-3.298455}--\num{-0.270739}] \\
  ZKsync    & \num{0.000884} & \num{0.000001}  & \num{-0.441471} & [\num{-0.658738}--\num{-0.255840}] \\
  \bottomrule
\end{tabular}
\end{table}

\begin{figure}[btp]
 % \centering
  \makebox[\textwidth][c]{%
  \includegraphics[width=1.2\linewidth]{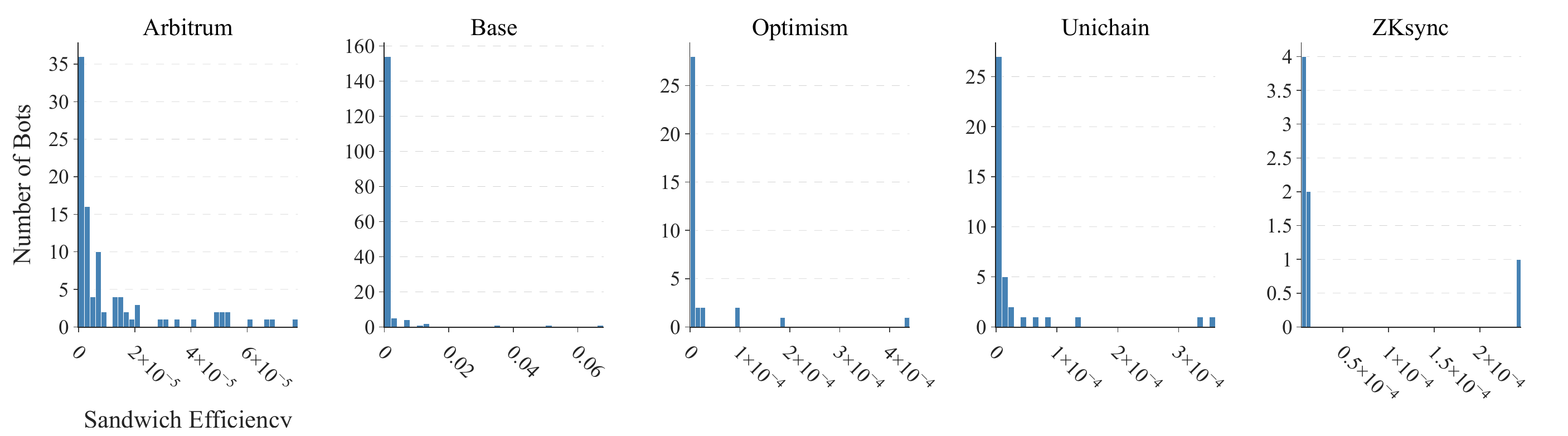}}
    \caption{Histograms of sandwich efficiency per bot across L2s (Jan--Sep'25). Efficiency is defined as the share of sandwiches among a bot’s total transactions. Across all rollups, the vast majority of bots exhibit extremely low efficiency, indicating that most sandwich-like patterns are not genuine MEV attacks.}
  \label{fig:sandwich-efficiency}
\end{figure}

% JG: moved to better ustilize a float-only page.
%\begin{figure}[t]
%\centering
%  \includegraphics[width=\linewidth]{figures/violin_net_pnl.pdf}
% % \fbox{\rule[0pt]{0pt}{2in} \hspace{3in}} % placeholder box
%  \caption{Violin distribution of estimated net PnL in USD. The distributions are tightly centered around zero, with heavy tails, confirming that median profits are negligible and most candidate sandwiches are not profitable.}
 % \label{fig:violin-pnl}
%\end{figure}

\subsection{Bot Activity Characteristics}

Table~\ref{tab:sandwich-summary} summarizes the characteristics of addresses flagged as sandwich attackers across the studied L2s. For each rollup we report the number of distinct attackers, the total number of sandwich attempts observed during the Jan--Sep'25 window, and the per-attacker distributions of sandwiches and daily transactions. The data reveal stark heterogeneity: while ZKsync shows only $7$ addresses with a combined $27$ sandwiches, Arbitrum and Base each host dozens to hundreds of bots with thousands of attempts. Transaction activity is equally skewed, with daily averages ranging from a few dozen to several hundred thousand per attacker. Outliers such as a single Arbitrum bot with $382{,}739$ transactions per day or a Base bot with $11{,}614$ sandwiches dominate aggregate counts, indicating that sandwiching is highly concentrated.

\begin{table}[tbp]
\centering
\sisetup{
  round-mode=places,
  round-precision=0,
  scientific-notation=false,
  group-separator={,}
}
\caption{Per-chain activity of bot addresses flagged as sandwich initiators. The large number of daily transactions compared to the relatively low cumulative number of sandwiches per bot suggests that most cases are accidental by-products of broader bot activity. Sandwich counts and averages are reported per bot.}
\label{tab:sandwich-summary}
\begin{tabular}{l r | rrr | rrr}
  \toprule
    Chain & \# Bots  
        & \multicolumn{3}{c|}{Sandwiches (Jan--Sep’25)} 
        & \multicolumn{3}{c}{Avg.\ Daily Tx} \\
  %\cmidrule(lr){3-5} \cmidrule(lr){6-8}
  %Chain & \# Bots & \multicolumn{3}{c|}{per Bot} 
  %   & \multicolumn{3}{c}{per Bot} \\
  \cmidrule(lr){3-5} \cmidrule(lr){6-8}
    &  & Min & Median & Max & Min & Median & Max \\
  \midrule
  Arbitrum & 96  & 1 & 5  & 321    & 55  & 5,968  & 382,739 \\
  Base     & 169 & 1 & 14 & 11,614 & 72  & 6,315  & 2,465,363 \\
  Optimism & 36  & 1 & 3  & 37     & 19  & 5,324  & 223,562 \\
  Unichain & 40  & 1 & 6  & 175    & 649 & 3,778  & 248,500 \\
  ZKsync   & 7   & 1 & 1  & 15     & 334 & 609    & 1,434 \\
  \bottomrule
\end{tabular}
\end{table}

To further examine how specialized these bots are, we measure \emph{sandwich efficiency}, defined as the share of sandwiches among a bot’s total transactions. Figure~\ref{fig:sandwich-efficiency} shows the histogram of bots' efficiency. In all chains, the median efficiency remains extremely low---often below $10^{-3}$---with only a few long-tailed cases of concentrated sandwiching. This suggests that the majority of addresses identified by heuristics are not dedicated sandwichers, but instead engage in broader arbitrage or market-making strategies, with sandwiches representing a marginal fraction of their flow. The joint evidence from Table~\ref{tab:sandwich-summary} and Figure~\ref{fig:sandwich-efficiency} underscores that sandwich attacks on L2s are rare, sporadic, and typically overshadowed by other forms of trading activity.

% JG: moved to utilize float-page
%\begin{figure}[t]
% % \centering
%  \makebox[\textwidth][c]{%
%  \includegraphics[width=1.2\linewidth]{figures/efficiency_histograms.pdf}}
%    \caption{Histograms of sandwich efficiency per bot across L2s (Jan--Sep'25). Efficiency is defined as the share of sandwiches among a bot’s total transactions. Across all rollups, the vast majority of bots exhibit extremely low efficiency, indicating that most sandwich-like patterns are not genuine MEV attacks.}
%  \label{fig:sandwich-efficiency}
%\end{figure}

%% file: sections/08ConclusionDiscussion.tex
\section{Discussion and Conclusions}
\label{sec:conclusions}

Sandwich attacks on Ethereum L1 have been extensively studied and are known to 
be both common and profitable. Their success relies on three structural 
features: deep liquidity that supports large victim trades, open builder markets 
that guarantee atomic inclusion of attacker legs, and transparent mempools that 
allow precise ordering control. These conditions collectively support a robust 
ecosystem of searchers who consistently extract value through sandwiches.

Our measurements show that none of these enabling conditions hold on L2 rollups.
Victim trades are an order of magnitude smaller, attacker legs fail to scale 
with victim size, and nearly all candidate patterns are economically 
inconsistent once slippage and fees are accounted for. Even when an attacker 
guesses the victim's pool and routing path correctly, centralized sequencing and 
private mempools eliminate atomic guarantees and reduce execution to a 
probabilistic event. The low co-inclusion probability, combined with shallow 
liquidity and narrow tick ranges, sharply limits feasible returns.

These findings have two key implications. First, heuristic detection methods may
substantially overestimate the prevalence of sandwiches on rollups by 
conflating innocent arbitrage with genuine victim-targeted attacks. Careful 
interpretation is required when labeling sandwich-like patterns in private 
mempool environments. Second, L2 defense mechanisms should not inherit L1-style 
assumptions: mitigation strategies must distinguish victim-targeted ordering 
manipulation from benign arbitrage flow, and policy discussions should reflect 
the fundamentally different sequencing model used by rollups.

\vspace{0.5em}
\noindent\textit{Limitations.}
Our analysis focuses on same-block patterns; on rollups with sub-second block
times, cross-block sandwiches are theoretically possible, though empirically 
rare. Actor attribution relies on transaction-tree heuristics and may merge 
unrelated flows, particularly when aggregator contracts bundle multiple hops.
Finally, our dataset covers the major rollups but may miss smaller venues or 
pools with different characteristics.

\vspace{0.5em}
\noindent\textit{Conclusions.}
This paper provides the first systematic analysis of sandwich attacks under 
private L2 mempools. We find that while the theoretical structure of sandwich 
profitability extends to rollups, practical execution does not: liquidity depth, 
slippage bounds, and low co-inclusion probability sharply limit feasible gains. 
Empirically, approximately 95\% of sandwich-shaped triples fail basic economic 
consistency checks, median profitability is negative, and attacker addresses 
exhibit negligible sandwich efficiency.

These results challenge the common assumption that sandwich behavior on 
Ethereum L1 generalizes to L2 environments. In private mempools controlled by 
centralized sequencers, sandwiches are rare, unprofitable, and often 
indistinguishable from neutral arbitrage. As L2 sequencing markets evolve—particularly if 
decentralized builders or preconfirmation markets emerge—the feasibility 
frontier for sandwiching may shift. Future work should monitor these changes, 
expand measurements to additional rollups, and explore cross-block patterns or 
searcher strategies that lie outside the classical same-block sandwich model.

%% file: sections/96Appendix.tex
\appendix
\section{Derivations and Proofs}
\label{app:proofs}

This appendix provides the formal derivations supporting the results of 
Section~\ref{sec:eco-model}. We first present the CPMM small-trade expansion 
leading to Proposition~1, then extend the argument to CLMMs and prove 
Proposition~2. 

\paragraph{Notation.} Throughout, we use the following convention:

\begin{itemize}
    \item $V_f$ : attacker frontrun input (USD),
    \item $V_v$ : victim input (USD),
    \item $\phi$ : proportional swap fee,
    \item $L$ : local liquidity depth (USD), 
    
 - for CPMM: $L = x_0$, the X-reserve;
 
 - for CLMM: $L = L_i/P$, tick liquidity normalized by price,

    \item $P$ : current price (Y per X),
    \item $[P_i, P_{i+1}]$ : CLMM ticks with constant liquidity $L_i$,
    \item All profits are reported in USD; intermediate CPMM expansions use 
          token units but convert back to USD via the small-trade approximation.
\end{itemize}

\subsection{CPMM Small-Trade Expansion (Proof of Proposition~1)}
\label{app:cpmm}

Consider a constant-product AMM with reserves $(x_0,y_0)$ and invariant
$k = x_0 y_0$. A trade of $\Delta X$ (X-input) with proportional fee $\phi$ has 
effective input $(1-\phi)\Delta X$ and output
\begin{equation}
\label{eq:cpmm-yout-app}
\Delta Y
=
\frac{y_0(1-\phi)\Delta X}{x_0 + (1-\phi)\Delta X}.
\end{equation}

\noindent
Define the normalized size
\[
\alpha = \frac{(1-\phi)\Delta X}{x_0},
\qquad \alpha \ll 1.
\]
Using the standard expansion 
$\frac{\alpha}{1+\alpha} = \alpha - \alpha^2 + O(\alpha^3)$,
we obtain
\begin{equation}
\label{eq:cpmm-expansion}
\Delta Y
=
\frac{y_0}{x_0}(1-\phi)\Delta X
\;-\;
\frac{y_0}{x_0^2}(1-\phi)^2 (\Delta X)^2
+
O(\Delta X^3).
\end{equation}

\paragraph{Frontrun, victim, backrun sequence.}

Let $V_f$ and $V_v$ be attacker and victim X-inputs.  
The frontrun moves reserves to  
$x_1 = x_0 + (1-\phi)V_f$;  
the victim trade moves to  
$x_2 = x_1 + (1-\phi)V_v$.

\noindent
Using symmetry of the CPMM invariant, the backrun output admits the expansion
\begin{equation}
\label{eq:xb-app}
\Delta X_b
=
(1-\phi)^2 V_f
+
\frac{(1-\phi)^2}{x_0}
\left(V_f V_v - V_f^2\right)
+
O\!\left( \frac{(V_f + V_v)^3}{x_0^2} \right).
\end{equation}

\paragraph{Incremental profit.}

Let $\Pi(V_f;V_v)$ denote profit with victim present and $\Pi(V_f;0)$ the 
baseline without a victim. Their difference eliminates attacker-only effects:
\[
\Delta \Pi(V_f;V_v)
=
\Pi(V_f;V_v) - \Pi(V_f;0).
\]
Substituting~\eqref{eq:xb-app} and simplifying yields the quadratic form
\begin{equation}
\label{eq:quad-app}
\Delta \Pi(V_f;V_v)
=
\frac{(1-\phi)^2}{x_0}
\left(V_f V_v - V_f^2\right)
+
O\!\left(\frac{(V_f + V_v)^3}{x_0^2}\right).
\end{equation}
Since in Section~2.1 we define $L = x_0$, this matches  
\[
\Delta\Pi \approx \frac{(1-\phi)^2}{L}(V_f V_v - V_f^2).
\]

\paragraph{Proof of Proposition~1.}

The leading-order profit is a strictly concave quadratic in $V_f$.  
Maximizing~\eqref{eq:quad-app} gives the interior optimum
\[
V_f^\star = \frac{V_v}{2}.
\]
Imposing the victim’s slippage constraint $0 \le V_f \le V_f^{\max}$ yields the 
global solution
\[
V_f^\star = \min\{V_v/2,\, V_f^{\max}\}.
\qquad\blacksquare
\]

\bigskip

% =============================================================
\subsection{CLMM Tick-Level Analysis (Proof of Proposition~2)}
\label{app:clmm}

A CLMM partitions the price axis into ticks $[P_i, P_{i+1}]$, each with 
constant liquidity $L_i$. The square-root price satisfies 
$\sqrt{P}\in[\sqrt{P_i},\sqrt{P_{i+1}}]$ inside the tick.  
Crossing a tick consumes
\begin{align}
\Delta y &= L_i(\sqrt{P_{i+1}} - \sqrt{P_i}),\\
\Delta x &= L_i\left(\frac{1}{\sqrt{P_i}} - \frac{1}{\sqrt{P_{i+1}}}\right).
\end{align}

\noindent
Inside the tick, the local depth is
\[
L = \frac{L_i}{P},
\]
and the CPMM small-trade expansion applies exactly.  
Thus, if $V_f + V_v$ stays within $[P_i,P_{i+1}]$:
\[
\Delta\Pi(V_f;V_v)
\approx
\frac{(1-\phi)^2}{L}
\left(V_f V_v - V_f^2\right),
\]
with optimum $V_f^\star = V_v/2$ (capped by slippage).

\paragraph{Boundary effects.}

At $P_{i+1}$ liquidity changes from $L_i$ to $L_{i+1}$.  
The marginal price impact with respect to X-input inside a tick is
\begin{equation}
\label{eq:dpdx-app}
\frac{dP}{dx}
=
\frac{2P^{3/2}}{L_i}.
\end{equation}

If $L_{i+1} < L_i$, then $1/L_{i+1} > 1/L_i$ and the slope~\eqref{eq:dpdx-app}
\emph{increases} upon entering tick $i{+}1$.  
Thus, the victim’s remaining input in the thinner tick produces strictly larger 
price movement, increasing the attacker’s round-trip wedge.

Let $V_f^{\mathrm{gap}}$ denote the minimal frontrun needed to reach $P_{i+1}$.

\paragraph{Proof of Proposition~2.}

\begin{itemize}
    \item \textit{Case 1: Remaining inside tick $i$.}  
    The profit is a concave quadratic with unique maximizer 
    $V_f^\star = V_v/2$.

    \item \textit{Case 2: Crossing into tick $i+1$ with $L_{i+1} < L_i$.}  
    Using~\eqref{eq:dpdx-app}, the marginal price impact jumps upward at 
    $P_{i+1}$, implying
    \[
    \Pi_{\mathrm{cross}}(V_f^{\mathrm{gap}}) 
    \;>\; 
    \Pi_{\mathrm{in}}(V_v/2).
    \]
    Hence the attacker strictly benefits from nudging the victim into the 
    lower-liquidity region.  
    The global optimum is
    \[
    V_f^\star = \min\{V_f^{\mathrm{gap}},\, V_f^{\max}\}.
    \]

    \item \textit{Case 3: $L_{i+1} \ge L_i$.}  
    No upward jump in~\eqref{eq:dpdx-app} occurs.  
    Boundary crossing does not increase profit, so the interior solution 
    $V_f^\star = V_v/2$ remains optimal.
\end{itemize}